\documentclass{aa} 
\usepackage{natbib}

\usepackage{mathrsfs}
\usepackage{bbm}
\usepackage{bm}
\usepackage{dsfont}
\usepackage{tensor}
\usepackage{xspace}
\usepackage{wasysym}
\usepackage{microtype}
\usepackage{empheq}
\usepackage{ifthen}
\usepackage{amsmath}
\usepackage{wrapfig}
\usepackage[colorlinks=true,linkcolor=blue,citecolor=blue, urlcolor=blue]{hyperref}
\usepackage[capitalise]{cleveref}
\usepackage{xcolor}
\usepackage{siunitx}
\DeclareSIUnit\parsec{pc}
\usepackage{floatrow}
\usepackage[normalem]{ulem}
\newcommand\rd{{\rm d}}

\newcommand\bbeta{\bm{\beta}}
\newcommand\btheta{\bm{\theta}}

\newcommand{\med}{\mathrm{med}}

\newcommand\e[1]{_{\mathrm{#1}}}
\newcommand\h[1]{^{\mathrm{#1}}}

\newcommand{\delimiters}[4][]{
\ifthenelse{ \equal{#1}{1} }{  #2 #3 #4  }
					{ \ifthenelse{\equal{#1}{2}}{ \big#2 #3 \big#4 }
						{ \ifthenelse{\equal{#1}{3}}{ \Big#2 #3 \Big#4 }
							{ \ifthenelse{\equal{#1}{4}}{ \bigg#2 #3 \bigg#4 }
								{ \ifthenelse{\equal{#1}{5}}{ \Bigg#2 #3 \Bigg#4 }
									{ \left#2 #3 \right#4 }
								}
							}
						}
					}
													}
													
\newcommand{\pa}[2][]{\delimiters[#1]{(}{#2}{)}}
\newcommand{\pac}[2][]{\delimiters[#1]{[}{#2}{]}}

\definecolor{blue4}{RGB}{0,0,143}
\definecolor{red4}{RGB}{143,0,0}
\definecolor{orange}{RGB}{255,128,0}
\definecolor{darkcyan}{RGB}{0,128,128}
\definecolor{olive}{RGB}{0,128,0}
\definecolor{purple}{RGB}{128,0,128}
\definecolor{cyan2}{RGB}{0,255,255}
\definecolor{fushia}{RGB}{255,0,255}
\definecolor{mygray}{gray}{0.5}
\definecolor{lightgray}{gray}{0.85}

\usepackage{graphicx}	
\usepackage{xcolor}     
\usepackage{soul}       
\usepackage{array}
\usepackage{bm}
\bibpunct{(}{)}{;}{a}{}{,}
\usepackage[OT1]{fontenc}
\usepackage{orcidlink}

\usepackage{newtxtext,newtxmath}

\begin{document}


\title{{\Large TDCOSMO XXXI.}\\ New techniques in line-of-sight studies of time delay lenses}

\author{
D.~P.~Johnson\inst{\ref{lupm}}\thanks{Corresponding authors: \href{mailto:daniel.johnson@umontpellier.fr}{daniel.johnson@umontpellier.fr}}\orcidlink{0000-0002-0311-2513}
\and
D.~Williams \inst{\ref{ucla}}\orcidlink{0000-0002-8386-0051}
\and
C.~D.~Fassnacht \inst{\ref{UCdavis}}\orcidlink{0000-0002-4030-5461}
\and
P.~R.~Wells \inst{\ref{argonne}}\orcidlink{0000-0003-0999-2395}
\and
P.~Fleury \inst{\ref{lupm}}\orcidlink{0000-0001-9292-3651}
\and
A.~G.~Schweinfurth \inst{\ref{tum},\ref{mpa}}\orcidlink{0000-0002-8274-7196}
\and
M.~Millon \inst{\ref{unigeth}}\orcidlink{0000-0001-7051-497X}
\and
A.~Galan \inst{\ref{unige}}\orcidlink{0000-0003-2547-9815}
}

\institute{
Laboratoire Univers et Particules de Montpellier (LUPM), CNRS \& Universit\'{e} de Montpellier (UMR-5299), Parvis Alexander Grothendieck, F-34095 Montpellier Cedex 05, France
 \label{lupm}
\goodbreak
\and
Department of Physics and Astronomy, University of California, Los Angeles, CA 90095, USA
\label{ucla}
\goodbreak
\and
Department of Physics and Astronomy, University of California, Davis, 1 Shields Ave., Davis, CA 95616, USA
 \label{UCdavis}
\goodbreak
\and
High Energy Physics Division, Argonne National Laboratory, Lemont, IL 60439, USA
\label{argonne}
\goodbreak
\and
Technical University of Munich, TUM School of Natural Sciences, Department of Physics, James-Franck-Str. 1, 85748 Garching, Germany 
\label{tum} 
\goodbreak
\and
Max-Planck-Institut für Astrophysik, Karl-Schwarzschild-Str. 1, 85748 Garching, Germany 
\label{mpa}
\goodbreak
\and
D\'epartement de Physique Th\'eorique, Universit\'e de Gen\`eve, 24 quai Ernest-Ansermet, CH-1211 Gen\`eve 4, Switzerland
\label{unigeth}
\goodbreak
\and
Department of Astronomy, University of Geneva, ch. d’Ecogia 16, 1290 Versoix, Switzerland 
\label{unige}
}

\abstract{
The distribution of matter along the same line of sight but external to the main lens, quantified by $\kappa\e{ext}$, is a key source of uncertainty in time delay cosmography and other applications of strong lensing, and must be independently 
estimated to avoid biasing the inferred value of the Hubble constant ($H_0$). 
We present advancements and standardisations in the weighted number counts techniques used to constrain $\kappa\e{ext}$, in particular the use of the \textit{Euclid} Flagship Simulation and a breakdown of the line-of-sight contributions between the observer, lens and source. As part of the TDCOSMO 2026 milestone analysis, we apply this updated method to the sample of 11 time delay lenses used in that study. Our estimates for certain systems are sensitive to these methodological changes, but are nonetheless consistent within $1\sigma$ for all but one of the systems which had been studied previously, with the median $\kappa\e{ext}$ across those systems changing from $-0.002$ to $-0.006$ following this new analysis. This work represents the first estimate of $\kappa\e{ext}$ values which includes the contribution of the observer-lens and lens-source terms, and largest standardised analysis of time delay lens environments to date.}


\titlerunning{New techniques in environmental studies}

\maketitle

\section{Introduction}
Over the last decade, the so-called ``Hubble tension'', a $\sim5\sigma$ discrepancy between the measured value of the Hubble Constant $H_0$ from Type-1a supernovae calibrated with Cepheid variable stars \citep{Riess_2022}, and the model-dependent value inferred from the CMB \citep{Planck_2020}, has proved a persistent challenge to paradigmatic cosmological models \citep{Di_Valentino_2025}. Before meaningful theoretical conclusions can be drawn, independent probes of $H_0$ are needed to shed light on the origin of this tension.

Time delay cosmography presents one such alternative. The relative time delays between the multiple images of a compact, time varying source depend both on the projected mass distribution between the observer and source, and on the background cosmology via a ratio of angular diameter distances which is particularly sensitive to $H_0$ \citep{Refsdal1964}. High resolution observations of lensed images constrain this mass distribution, and so, in conjunction with high cadence observations of the image fluxes and hence measurements of the relative time delays between them, can be used to place constraints on $H_0$ which are independent of both distance ladder methods and the CMB \citep[e.g.,][]{birrer_2024}. The TDCOSMO collaboration uses a population of galaxy-scale lenses with variable quasar sources to hierarchically infer $H_0$ and other cosmological parameters via this technique, with the goal of achieving sufficient precision to select between the values preferred by early and late-universe probes of $H_0$ \citep{2025A&A...704A..63T}. 

Strong gravitational lensing is fundamentally sensitive to the path taken by a light beam between the observer and source. This path is determined by both the distribution of matter in the main lens galaxy, as well as that lying on or near the line of sight. Any smooth over or under-density of matter along the lens line of sight, quantified via the external convergence, $\kappa\e{ext}$, mimics a change to the background cosmology in the impact it has on light propagation, and strong lensing images and time delays alone cannot distinguish between them. This effect is known as the (external) mass-sheet degeneracy, and represents a key source of uncertainty in time delay cosmography. $H_0$ is directly degenerate with $\kappa\e{ext}$, and any net over or under density on time delay lens lines of sight will systematically bias the inferred value of $H_0$ unless corrected for \citep{Falco_1985}.

One route to correcting for the external contribution would be to explicitly model all visible structures along the line of sight. For large galaxies that are close in projection to the lens system, this explicit modeling is often necessary (\citealp{McCully2017}, see also \citealp{Li_2021}), and such objects are routinely included in lens models \citep[e.g.][]{Rusu2019, Birrer2019,Williams_2025,Paic_2026,2026arXiv260414145S, williams_2026}. However, any unmodelled (or incorrectly modelled) mass along the line of sight will, at first order, contribute to $\kappa\e{ext}$. Without extensive spectroscopic coverage and without strong assumptions about the mass structures of the haloes containing observed galaxies, a comprehensive accounting for the integrated line-of-sight contribution is impossible. 

Weak lensing measurements in the field in which the lens resides, which are sensitive to the entire projected matter distribution, can be used to reconstruct this convergence, and this technique has been used in \cite{Tihhonova_2018,Tihhonova_2020}. This approach, however, requires extremely high quality data to obtain precise measurements of the shapes of many high redshift galaxies, and obtaining convergence estimates from this data is highly non-trivial. 

The most frequently employed method to estimate $\kappa\e{ext}$, and the focus of this paper, is the `weighted number counts' method. The essential principle is that, while imperfect tracers of the total matter distribution, galaxies nonetheless carry information about the relative over or under density of a given line of sight. The idea is therefore to weight simulated lines of sight and their $\kappa\e{ext}$ values by how similar these are to the actual lens line of sight, according to the observable properties of the galaxies near these lines of sight. This weighted sample defines the probability distribution function (PDF) of the $\kappa\e{ext}$ value for that lens. The stronger the correlations between the chosen properties of the galaxy distribution and the true $\kappa\e{ext}$ value on that line of sight, the tighter this PDF will be.
The simulation used need not have the angular and mass resolution necessary to resolve strong lenses, nor to perfectly mimic the real line of sight of the observed lens. Provided that the relationship between the simulated galaxy catalog and the dark matter halos from which the convergence is determined is well calibrated statistically, by sampling many simulated sight lines weighted by the observed distribution of galaxies, the resulting $\kappa\e{ext}$ posterior is effectively marginalised over the uncertainties in that relationship.

Relative galaxy number counts was first introduced as a tracer of $\kappa\e{ext}$ in \cite{Suyu_2010,Fassnacht_2010}, and in \cite{Suyu_2013} the external shear values measured from lenses were first included as an additional weight by which simulated lines of sight were selected. In \cite{Greene_2013}, further weights were introduced, exploiting information from galaxy redshifts, angular separations, stellar mass and luminosity of galaxies in the lens field. The method has subsequently undergone a number of additional evolutionary steps, notably in \cite{Rusu_2017} and \cite{Birrer2020_TDC4}, optimising the trade-off between maximising information and the observational cost of obtaining it. In recent works, the method has been further standardised, and the simultaneous analysis of a larger sample of lenses has been made possible by the introduction of the python modules \texttt{lenskappa}\footnote{\href{https://github.com/AstroPatty/cosmap}{https://github.com/AstroPatty/cosmap}} and \texttt{cosmap}\footnote{\href{https://github.com/AstroPatty/lenskappa}{https://github.com/AstroPatty/lenskappa}} \citep{Wells_2023,Wells_2024}.

The weighted number counts method has been shown to yield estimates of the observer-to-source convergence $\kappa\e{s}$ which are consistent with weak lensing estimates, justifying its use as a much simpler analysis with only a marginal loss of precision \citep{Rusu_2017,Tihhonova_2018}. However, the quantity with which $H_0$ is degenerate is not simply the $\kappa\e{s}$ obtained from simulations, but a combination of this term with the observer-lens and lens-source terms $\kappa\e{d}$ and $\kappa\e{ds}$, the latter of which is less straightforward to obtain from simulation outputs. Even at leading order, the contribution of these terms to $\kappa\e{ext}$ may not be negligible, and their omission when using additional constraints from velocity dispersion measurements can also introduce a bias of $\kappa\e{d}$ on the inferred value of $H_0$ \citep{Teodori_2022}.

In this work, we introduce updates to the weighted number counts method, which enable $\kappa\e{d}$, $\kappa\e{ds}$ and $\kappa\e{s}$ to be constrained individually. We argue that this change makes estimates of $\kappa\e{ext}$ more accurate. We update \texttt{cosmap} to accommodate the \textit{Euclid} Flagship Simulation \citep{Castander_2025}, which we argue offers a number of advantages over the Millennium Simulation \citep{Springel_2005} on which previous analyses relied. We argue against the use of the external shear as a source of additional constraint. 

In addition, we apply the weighted number counts analysis with our updates to the TDCOSMO 2026 milestone sample of 11 time delay lenses, and these results are used in the milestone analysis (TDCOSMO 2026, in prep.). Our analysis was ``blind'' with respect to the median values of our $\kappa\e{ext}$ posteriors, and no changes were made to the results presented in this work after they were unblinded. While certain lenses individually lie on over or under dense lines of sight, there is no obvious systematic trend.

In \cref{sec:theory}, we present the theoretical background to this work - the lens and time delay equations, the role of the line-of-sight mass distribution, and the resulting impact on $H_0$ inference. In \cref{sec:number_counts}, we summarise the weighted number counts method, and highlight our changes and their motivations in \cref{sec:updates}. In \cref{sec:application}, we apply this method to our time delay lens sample. We present our conclusions in \cref{sec:conclusion}.

\section{Theoretical background \label{sec:theory}}

\subsection{Lensing equation and line-of-sight}

If light from a source is lensed by a main lens with potential $\psi(\btheta)$ in the presence of tidal line-of-sight perturbations, the source position $\bbeta$ will be related to the position $\btheta$ at which the light is observed via
\begin{equation}
    \bbeta = \left(1-\kappa\e{s}\right) \btheta - \left(1-\kappa\e{ds}\right)
    \frac{ \rd}{\rd \btheta}
     \psi\left[\left(1-\kappa\e{d}\right) \btheta\right], \label{eq:lens_equation}
\end{equation}
where $\kappa\e{s}$, $\kappa\e{d}$ and $\kappa\e{ds}$ are the tidal convergence terms between observer and source, observer and deflector, and deflector and source respectively. For simplicity we omit the line-of-sight shear terms, but a full treatment can be found in e.g. \cite{Fleury_2021a}. 

The convergence $\kappa_{ij}$ between two planes at comoving distances $\chi_i<\chi_j$ is a dimensionless measure of the projected surface mass overdensity relative to the average density of the universe (such that a negative $\kappa_{ij}$ corresponds to an underdensity between those planes). 
From \cite{Fleury_2019a}, $\kappa_{ij}$ can be written, for a spatially flat universe and at linear order in the matter density contrast $\delta$ at conformal time $\eta$, as a weighted integral of $\delta$,
\begin{equation}
    \kappa_{ij} = \frac{3}{2}\Omega\e{m0}H_0^2\int^{\chi_j}_{\chi_i}\rd \chi \; (1+z)K_{ij}(\chi)\delta[\eta_0-\chi,\chi,\chi\btheta],
    \label{eq:kappa_ij_delta}
\end{equation}
where $\eta_0$ is the conformal time today, $\Omega\e{m0}$ is the matter density parameter today, and
\begin{equation}
    K_{ij}(\chi) \equiv \frac{(\chi_j-\chi)(\chi-\chi_i)}{\chi_j-\chi_i}. \label{eq:weight_function}
\end{equation}

The source position $\bbeta$ is never directly observed, and thus, following e.g. \cite{Birrer2020_TDC4,Fleury_2021a}, \cref{eq:lens_equation} can be rewritten as 
\begin{equation}
    \tilde{\bbeta} = (1-\kappa\e{ext})\btheta - 
    \frac{\rd \psi\e{eff}(\btheta)}{\rd \btheta},
    \label{eq:minimal_model}
\end{equation}
with the transformed source position $\tilde{\bbeta} \equiv (1-\kappa\e{d})(1-\kappa\e{ds})^{-1}\bm{\beta}$, the effective gravitational potential $\psi\e{eff}(\bm{\theta}) \equiv \psi[(1-\kappa\e{d})\bm{\theta}]$, and the ``external'' (or ``line-of-sight'') convergence  
\begin{equation}
    1-\kappa\e{ext} \equiv \frac{(1-\kappa\e{d})(1-\kappa\e{s})}{1-\kappa\e{ds}} \approx 1 - \kappa\e{d}-\kappa\e{s} + \kappa\e{ds}.
    \label{eq:kappa_ext_breakdown}
\end{equation}
\subsection{Time delays}
The time delay experienced by a strongly lensed light ray in the presence of tidal line-of-sight perturbers relative to an unlensed ray is given by
\begin{equation}
    t(\btheta,\bbeta) = \frac{D_{\Delta t}}{c}\phi(\btheta,\bbeta),
    \label{eq:time_delay_tidal_general}
\end{equation}
where the (background) time delay distance $D_{\Delta t}$ \citep{Suyu_2010} and the Fermat potential $\phi(\btheta,\bbeta)$ are defined by
\begin{gather}
    D_{\Delta t} \equiv (1+z_{\rm d})\frac{D\e{d}D\e{s}}{D\e{ds}},
    \label{eq:time_delay_scale} \\
    \phi(\btheta,\bbeta) \equiv \frac{1}{2}(1-\kappa\e{ext})(\btheta-\bbeta')^2-\psi_{{\rm eff}}(\btheta), \label{eq:Fermat_potential}
\end{gather}
and where $\bbeta' \equiv (1-\kappa\e{s})^{-1}\bbeta$. The quantity $t(\btheta,\bbeta)$ is unmeasurable, but relative time delays $\Delta t$ between lensed images of the same source are. Such time delays, via their dependence on $D_{\Delta t}$, are inversely proportional to the Hubble constant,
\begin{equation}
    \Delta t \propto \frac{1}{H_0}.
\end{equation}
Thus, if the lensing potential $\psi\e{eff}$ and line-of-sight contribution $\kappa\e{ext}$ can be constrained, measurements of image positions $\btheta$ and the time delays between them allow constraints to be placed on $H_0$.

\subsection{The external mass-sheet degeneracy}

Suppose \cref{eq:minimal_model} is divided throughout by $(1-\kappa\e{ext})$, to obtain
\begin{equation}
    \bbeta' = \btheta - 
    \frac{\rd \psi'\e{eff}(\btheta)}{\rd \btheta},
    \label{eq:actual_model}
\end{equation}
where
\begin{gather}
    \psi'\e{eff} = (1-\kappa\e{ext})^{-1}\psi\e{eff}.
\end{gather}
The unobservability of the source position ensures that \cref{eq:actual_model} is equally applicable to any lensed image observations. The indistinguishability of \cref{eq:lens_equation,eq:actual_model} is known as the (external) mass-sheet degeneracy (MSD) \citep{Falco_1985,Schneider_2013,Birrer_2016,Kochanek_2020}. In practice, therefore, \cref{eq:actual_model} is typically used in lens modelling, with a lens potential $\psi'\e{eff}(\btheta)$ which captures the deflection of the main lens and the foreground convergence, up to a rescaling by an unknown $(1-\kappa\e{ext})$, to which an ``external'' (or ``residual'') shear term $\bm{\Gamma}\e{ext}$ is often added to account for both shear from matter along the line of sight and any leftover shear-like complexity in the lens galaxy not captured by $\psi'\e{eff}(\btheta)$ \citep{Shajib_2024}. 

If lens images are used to constrain $\psi'\e{eff}(\btheta)$, the Fermat potentials that will be predicted are 
\begin{align}
    \phi'(\btheta,\bbeta') &= \frac{1}{2}(\btheta-\bbeta')^2-\psi'_{{\rm eff}}(\btheta), \\
    &= (1-\kappa\e{ext})^{-1}\phi(\btheta,\bbeta'). 
\end{align}
If relative time delays are measured, a value $D'_{\Delta t} = (1-\kappa\e{ext})D_{\Delta t}$ will be inferred which compensates for the missing factor of $(1-\kappa\e{ext})$ in $\psi'_{{\rm eff}}(\btheta)$. As $D_{\Delta t} \propto 1/H_0$, the inferred value of the Hubble constant from $D'_{\Delta t}$, $H\h{mes}_0$, will be biased compared to the true value $H_0$ by a factor
\begin{equation}
    \frac{H\h{mes}_0}{H_0} = (1-\kappa\e{ext})^{-1}.
\end{equation}
The MSD makes a constraint of $H_0$ from lens images and relative time delay measurements impossible without independent constraints or priors on $\kappa\e{ext}$. 

\section{Estimating the external convergence with weighted number counts \label{sec:number_counts}}
The external convergence is a weighted projection of the density contrast between the observer and source which is not accounted for within the main lens and any perturbing galaxies explicitly included in the lens model/s. Although much of this mass is expected to be non-baryonic, the distribution of galaxies in the vicinity of the lens and along its line of sight nonetheless carries information about the density contrast in that direction of the sky. By comparing the relative over- or under-density of this galaxy distribution to simulated lines of sight in which the full mass distribution and hence convergence is known, the external convergence of a lens can be estimated. In the following, we summarize the weighted number counts method set out in \cite{Wells_2023,Wells_2024}, our choice of weights and analysis region, which perturbers are excluded from the counts, and our choices with regards to photometric redshifts. 
\begin{table*}[t]
\caption{Comparison between the Millennium Simulation \citep{Springel_2005} and the \textit{Euclid} Flagship FS2 \citep{Castander_2025}.}
\label{tab:sim_comparison}
\centering
\small
\renewcommand{\arraystretch}{1.3}
\begin{tabular}{lp{5.8cm}p{5.8cm}}
\hline\hline
& Millennium Simulation (MS, 2005) & \textit{Euclid} Flagship (FS2, 2024) \\
\hline
Volume / geometry &
  $500\,h^{-1}\mathrm{Mpc}$ ($0.125\,\mathrm{Gpc}^3$), periodic box &
  $3600\,h^{-1}\mathrm{Mpc}$ ($46.7\,\mathrm{Gpc}^3$), full-sky lightcone\\[2pt]
DM particles &
  $2160^3 \approx 1.0\times10^{10}$ &
  $16 000^3 \approx 4.0\times10^{12}$ \\[2pt]
Particle mass &
  $8.6\times10^8\,h^{-1}M_\odot$ &
  $1\times10^9\,h^{-1}M_\odot$ \\[2pt]
Ray-tracing angular resolution &
  Nside = 4096 ($\approx 0.86'$ pixel resolution) &
  Nside = 8912 ($\approx 0.4'$ pixel resolution)  \\[2pt]
Force softening &
  $5\,h^{-1}\mathrm{kpc}$ &
  $4.5\,h^{-1}\mathrm{kpc}$ \\[2pt]
Cosmology &
  $\Omega\e{m}=0.25$, $\Omega\e{b}=0.045$, $\sigma_8=0.90$, $n_s=1$, $h=0.73$ (2dF-like) &
  $\Omega_m=0.319$, $\Omega_b=0.049$, $\sigma_8=0.83$, $n_s~=~0.96$, $h=0.67$ (Planck-like \textit{Euclid} reference cosmology) \\[2pt]
Galaxy modelling &
  Semi-analytic \cite{Croton_2006}; &
  Halo occupation distribution and abudance matching \cite{Castander_2025} \\[2pt]
\hline
\end{tabular}
\end{table*}

\subsection{The weighted number counts method}
The core approach of the weighted number counts method is to use a set of summary statistics $\boldsymbol{W}$ of the distribution of galaxies close to the line of sight to the lens, determined relative to arbitrarily sampled directions in a reference survey, as a tracer of the relative over or under density of the lens line of sight. Galaxies contribute towards these statistics if:
\begin{enumerate}
    \item they fall below (i.e. are brighter than) a chosen magnitude threshold in the band of interest, to maximise their information content while still being well-covered by the lens environment observation and reference survey;
    \item they lie within an annulus around the lens, defined by minimum and maximum cutoff radii $r\e{min}$ (to exclude the lens itself) and $r\e{max}$ \citep[beyond which the contribution of galaxies to the environmental characterisation is taken to be negligible, see][]{Collett2013}
    \item they lie below the redshift of the source, if redshift information is available, as matter at $z>z\e{s}$ will not contribute to $\kappa\e{ext}$;
    \item they are not explicitly included in the model (see \cref{sec:modelled_perturbers}).
\end{enumerate}
Galaxies satisfying these conditions are used to determine a series of weights $w_j$, which quantify the contribution of galaxy $j$ to a corresponding summary statistic $W_i$, which is expected to correlate with the convergence in that field $i$. A range of weighting schemes have been explored in the literature \citep[see e.g.][]{Greene_2013,Rusu_2017}; in this work, we follow \cite{Wells_2023,Wells_2024} and calculate, for each galaxy $j$ within the sampled region,
\begin{gather}
    w_j^{(n)}=1 \quad \text{(pure number counts)}, \\
    w_j^{(1/r)} = 1/r_j \quad \text{(inverse distance weights)},  \\
    w_j^{(z/r)} = z_j(z\e{s}-z_j)/r_j \quad \text{(redshift--distance weights)},
\end{gather}
where $r_j$ is the angular distance between galaxy $j$ and the lens centre, and $z_j$ its redshift. If no redshift information is available, we omit $w_j^{(z/r)}$. 

For each field $i$ within the reference survey, the summary statistic $W_i$ is then calculated as 
\begin{equation}
W_i
\equiv
\frac{n\e{lens} \, \med\pa{\{w_j\}_{j\in \text{lens field}}}}
   {n_i \, \med\pa{\{w_j\}_{j\in \text{field $i$}}}} ,
\end{equation}
where $n_i$ is the number of galaxies in the field $i$, $n\e{lens}$ the number of galaxies in the lens field, and $\med\pa{...}$ is the median value of a given weight for those galaxies \citep{Rusu_2017}. The quantities $W_i$ thus capture information about the relative over or under-density of the lens field compared to the field $i$. This procedure is repeated for many reference fields using \texttt{cosmap}, and an $N$-dimensional histogram is built over the $N$ chosen weights ($N=3$ if redshift information is available, and $N=2$ if not). This histogram characterises the relative density of the lens field compared to a representative sample of arbitrary directions in the sky.

The same procedure is then used to generate corresponding summary statistics within a simulated survey of galaxies, across many simulated lines of sight. Previous studies made use of the Millennium Simulation~\citep{Springel_2005}; here we upgrade to the \textit{Euclid} Flagship Simulation \citep{Castander_2025} -- see \cref{sec:Euclid_flagship}. The weight $W_i$ for a given simulated line of sight~$i$ is normalised by the median of the corresponding weights across all the sampled simulated lines of sight, so that the simulated statistic has a median of unity, and these quantities once again characterise the relative density of a given line of sight compared to the population. These simulated lines of sight are then weighted according to the weight-space densities of the field counts, thus reweighting their distribution according to our knowledge of the the characteristics of the lens line of sight.\footnote{In this analysis, prior to unblinding our results, field counts in the 2.5 upper and lower percentile were omitted as outliers (extreme comparison fields not well captured by the simulation, problematic data etc). Tests suggest that this choice had a small impact on the resulting $\kappa\e{ext}$ constraints, with the median $\kappa\e{ext}$ and the width of the 68\% confidence interval shifting by $\sim 10\%$ percent of the 1-$\sigma$ uncertainty, with no preferential direction for this shift. This is something we intend to explore more in future analyses, to prevent a mischaracterisation of any physically meaningful posterior tail, but in this work we consider the effect to be negligible.} Associated with these lines of sight are the convergence outputs of the simulation, from which $\kappa\e{ext}$ can be determined (see \cref{sec:estimating_kappa_ext}). The resulting posterior $p(\kappa\e{ext}|\mathbf{d})$, the weighted distribution of $\kappa\e{ext}$ values given the observational data $\mathbf{d}$, is a marginalisation of the simulated $p(\kappa\e{ext}|\boldsymbol{W})$ over the observational constraint $p(\boldsymbol{W}|\mathbf{d})$,
\begin{equation}
    p(\kappa\e{ext}|\mathbf{d}) \propto \int p(\kappa\e{ext}|\boldsymbol{W})p(\boldsymbol{W}|\mathbf{d})\rd \boldsymbol{W}.
\end{equation}
\subsection{Modelled perturbers}
\label{sec:modelled_perturbers}
As the external convergence accounts for lensing by structures not included in the lens model itself, following previous analyses \citep[e.g.][]{Rusu2019, Birrer2019, 2026arXiv260414145S, williams_2026}
, a key step in generating number counts is to mask or remove from the lens environment catalog any strong perturbers whose flexion shift \citep[see][]{McCully2017} is sufficiently strong to necessitate their inclusion in the modelling. Defining the flexion shift as
\begin{equation}
    \Delta_3 = f(\beta\e{c}) \times \frac{\left(\theta\e{E}\theta\e{E,p}\right)^2}{\theta^3},
\end{equation}
where $\theta\e{E}$ and $\theta\e{E,p}$ are the Einstein radii of the main and pertuber lens galaxies respectively, $\theta$ their angular separation, and 
\begin{equation}
    f(\beta\e{c}) =     \begin{cases}
        (1-\beta\e{c})^2 & \mathrm{if} \quad z > z\e{d},\\
        1 & \mathrm{if} \quad z < z\e{d},
    \end{cases}
\end{equation}
where $\beta\e{c}$ is the cosmological scaling factor,
%
\begin{equation}
    \beta\e{c} = \frac{D\e{dp}D\e{s}}{D\e{p}D\e{ds}},
\end{equation}
and $D\e{p}$ and $D\e{dp}$ are the angular diameter distances to the perturber and from the deflector to the pertuber respectively, \cite{McCully2017} recommends treating any galaxy $\Delta_3x > 10^{-4}$arcsec as a strong perturber to be explicitly included within the lens model, and this criteria has been used in previous TDCOSMO analyses \citep[e.g.][for recent examples]{Paic_2026,2026arXiv260414145S}. Whatever the choice, any galaxy which is already accounted for within the lens model should be removed from the catalog to avoid ``double-counting'' its contribution to $\kappa\e{ext}$.
\subsection{Photometric redshifts}
In lens environment studies, spectroscopic redshifts are usually not complete to the depth needed for number count analyses. Instead, if  any comprehensive redshift estimates are present, these are typically photometric (as is the case for our lens sample), with non-negligible uncertainties. As a result, if these redshift estimates are compared directly to exact galaxy redshifts within a simulation, this uncertainty will be underestimated. When generating our redshift--distance weights in the simulation catalogue, we therefore make use of the \textit{Euclid}-like photometric redshift estimations provided in the FS2 simulation outputs, which introduces an additional scatter into the relationship between the apparent 3D distribution of galaxies and the corresponding convergence estimates.

\section{Updates to the method}
\label{sec:updates}
%

Weighted number counts have been used extensively in previous analyses of time delay lenses, but  theoretical and computational developments in the field provide an opportunity to revisit the current limitations of the method. The first of these is that the reliance on the Millennium simulation means that recent advancements in large cosmological simulations are not yet exploited. The second is the neglection of the foreground and background contributions to $\kappa\e{ext}$, which may not be negligible, and thus may introduce biases as other sources of uncertainty are more tightly controlled. Finally, analyses of certain time delay lenses have relied on the modelled external shear as an additional constraint, which is a potential source of systematic biases from mismodelled angular complexity. In this section, we present our three main changes with respect to the previous literature -- the use of the \textit{Euclid} Flagship Simulation, the breakdown of the line-of-sight contributions from $\kappa\e{d}$, $\kappa\e{s}$ and $\kappa\e{ds}$, and the exclusion of the external shear as a weight.
\subsection{The \textit{Euclid} Flagship simulation}
\label{sec:Euclid_flagship}
The first significant difference between our and previous analyses is the replacement of the Millennium Simulation (MS) \cite{Springel_2005} with the \textit{Euclid} Flagship Simulation (FS2) \citep{Castander_2025}, a massive N-body dark matter simulation populated with galaxies. The key relevant differences between these simulations are summarised in \cref{tab:sim_comparison}. While the two simulations use a comparable particle mass and softening scale, the FS2 offers a dramatic upgrade in simulation volume and statistical power, as well as better angular resolution, more up-to-date input cosmological parameters and more sophisticated galaxy populating models, via halo occupation distributions calibrated via abundance matching. The FS2 outputs needed for $\kappa\e{ext}$ estimation are directly available on CosmoHub \citep{Tallada_2020}\footnote{with the exception of the full weak lensing convergence and shear maps, which are available from the FS2 team on request for 200 redshift bins}, while the MS data is no longer available from any primary source. The FS2 outputs include galaxy magnitudes matching the properties of many different survey telescopes and their filters, limiting any errors which may arise from the post-processing necessary when using the MS.
\subsection{Estimating the external convergence}
\label{sec:estimating_kappa_ext}
\begin{figure}[!ht]
    \centering
    \includegraphics[width=\linewidth]{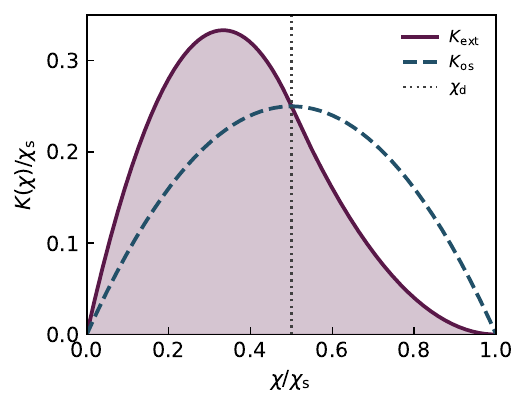}
    \caption{The observer-to-source weak lensing kernel $K\e{os}$ and the effective kernel $K\e{ext}$ of the external convergence $\kappa\e{ext}$, a combination of contributions from $\kappa\e{s}$, $\kappa\e{d}$ and $\kappa\e{ds}$, for a lens located at $\chi = 0.5\chi\e{s}$.}
    \label{fig:kernel_plot}
\end{figure}
The second key change featured in our analysis is the full calculation of $\kappa\e{ext}$ via the exact expression in \cref{eq:kappa_ext_breakdown}. Previous studies have assumed $\kappa\e{ext} \approx \kappa\e{s}$, which in turn requires that 
\begin{equation}
    \kappa\e{d} - \kappa\e{ds} \ll \kappa\e{s}.
\end{equation}
The conventional choice to use $\kappa\e{s}$ as a substitute for $\kappa\e{ext}$ was, in part, motivated by simplicity -- simulation outputs will typically include $\kappa\e{s}$ for a range of source redshifts, while $\kappa\e{ds}$ is not readily available. This choice, however, tends to under (over) estimate the contribution of foreground (background) objects, as shown in \cref{fig:kernel_plot}, and may skew the resulting distributions, especially if these foregrounds and backgrounds are strongly over or under-dense. 

In \cite{Johnson_2025}, it is shown that $\kappa\e{ds}$ can be determined directly from $\kappa(\chi)$, i.e. the observer-to-source convergence as a function of source comoving distance. Inverting \cref{eq:kappa_ij_delta}, $\delta$ can be written as a function of $\kappa\e{s}$, and thus
\begin{equation}
    \kappa_{ij} = \int^{\chi_j}_{\chi_i}\frac{\rd\chi}{\chi}\;W_{ij}\frac{\rd^2}{\rd\chi^2}\left[\chi\kappa(\chi)\right].
    \label{eq:kappa_ij_from_kappa_s}
\end{equation}
Thus, if $\kappa(\chi)$ is known along a given line of sight, $\kappa\e{s}$ and $\kappa\e{d}$ can be found be simply evaluating this function at the radial comoving distances to the source and lens, \cref{eq:kappa_ij_from_kappa_s} can be used to obtain $\kappa\e{ds}$, and $\kappa\e{ext}$ can be calculated without approximation from \cref{eq:kappa_ext_breakdown}.

Simulations such as FS2 typically produce convergence maps for $\kappa(z_n)$ evaluated on N redshift ``shells'' (200 between $z~\approx~0.003$ and $z~\approx~87$, in the case of FS2). In \cite{Johnson_2025}, $\kappa(\chi)$ was determined via a smooth interpolation of these $\kappa(z_n)$ over the corresponding values of $\chi_n$, and \cref{eq:kappa_ij_from_kappa_s} solved via a numerical integral. While accurate, this calculation inevitably adds a significant computational burden when many lines of sight are sampled. We therefore instead make use of a discrete integral estimate, the details of which can be found in Appendix \ref{app:discrete_integration}. 

The $\kappa\e{d}$ and $\kappa\e{s}$ values are determined via a linear interpolation between the shells adjacent to $z\e{d}$ and $z\e{s}$ respectively, and $\kappa\e{d}$, $\kappa\e{s}$, $\kappa\e{ds}$ and their combination as $\kappa\e{ext}$ via \cref{eq:kappa_ext_breakdown} are saved.
\subsection{The external shear}
\label{sec:external_shear}
When modelling lenses, a phenomenological ``external shear'', $\gamma\e{ext}$, is often included in the lens potential, without which models often fail to reproduce the lens image~\citep{1997ApJ...482..604K}. Nominally, this quantity accounts for the tidal shear arising from matter in the lens environment and along the line of sight \citep[the ``line-of-sight shear,'' $\gamma\e{LOS}$ in the language of][]{Fleury_2021a}. In theory, $\gamma\e{LOS}$ and $\kappa\e{ext}$ are closely connected, and previous studies have compared the measured $\gamma\e{ext}$ to values of $\gamma\e{s}$ (the observer-to-source weak lensing shear) within the simulated lines of sight as an additional constraint \citep[e.g.][]{Suyu_2010,Wells_2023}.

In practice, however, recent studies have demonstrated that measured values of $\gamma\e{ext}$ are almost certainly not entirely ``external'', but instead may contain contributions from the main lens itself \citep[see e.g.][]{etherington_2023}. In the language of \cite{Shajib_2024}, $\gamma\e{ext}$ should instead be thought of as a `residual shear', and is likely often a poor approximation for the magnitude of the weak lensing shear in that direction \citep{etherington_2023}. We therefore choose not to include any weights involving $\gamma\e{ext}$ in this analysis.
\section{Application to a sample of time-delay lenses \label{sec:application}}

\subsection{Lens sample and catalogue generation}
\begin{table*}[t]
\caption{The lenses studied in this work, their lens and source redshifts, the telescope/survey from which the observational data used to generate the lens environment catalogs was acquired, the work in which those catalogs were generated, and the source of the photometric redshifts used in the analysis, where available.}
\label{tab:lens_sample}
\centering
\small
\renewcommand{\arraystretch}{1.3}
\begin{tabular}{l p{3.0cm} p{3.0cm} p{5.0cm} p{2.5cm} }
\hline\hline
Lens &
Lens redshift &
Source redshift &
Lens environment observation and catalogue generation &
Photometric redshifts  \\
\hline
DES0408-5354 & 0.597 \cite{Lin_2017}  & 2.375 \cite{Lin_2017}  & DES\footnotemark[2]\cite{Abbott_2021} &  DES \cite{Abbott_2021}   \\
HE0435-1223  & 0.4546 \cite{morgan2005}  & 1.693 \cite{sluse2012} & Suprime-Cam \cite{Rusu_2017}  &  \cite{Rusu_2017}  \\
HE1104-1805  & 0.729 \cite{Lidman_2000} & 2.319 \cite{Lidman_2000}  &  Suprime-Cam, \cite{Paic_2026}  &   n/a   \\
PG1115+080  & 0.310 \cite{williams_2026}  & 1.727 (Knabel et al. 2026, in prep) & MPIA/ESO 2.2m \cite{Bonvin2018}, this work  &    n/a   \\
RXJ1131-1231 & 0.295 \cite{Sluse_2003}  & 0.657 \cite{Sluse_2007}   &  Suprime-Cam, this work   &    n/a \\
SDSS J1206+4332   & 0.745 \cite{Agnello_2015} & 1.789 \cite{Oguri_2005} &  WIYN\footnotemark[5], this work  &    n/a    \\
J1433+6007   & 0.407 \cite{Agnello_2018b}  & 2.737 \cite{Agnello_2018b} & DESI Legacy Imaging Survey \cite{Dey_2019, 2026arXiv260414145S}     &   \cite{Zhou_2023}  \\
J1537-3010   & 0.590 \cite{galan_2026} & 1.721 \cite{galan_2026}  & DELVE DR3\footnotemark[6] (Drlica-Wagner et al., in prep.), \cite{galan_2026} &  DECaLS DR10 \cite{Dey_2019}  \\
B1608+656   & 0.6304 \cite{Myers_1995} & 1.394 \cite{Fassnacht_1996}  &   Suprime-Cam, this work   &    n/a    \\
WFI2033-4723 & 0.6575 \cite{sluse2019} & 1.662 \cite{sluse2012} & DES \cite{Abbott_2021} &  DES \cite{Abbott_2021} \\
WGD2038-4008 & 0.2283 \cite{Buckley-Geer_2020} & 0.777 \cite{Agnello_2018} & DES \cite{Abbott_2021} &  DES \cite{Abbott_2021}  \\
\hline
\end{tabular}
\end{table*}
\Cref{tab:lens_sample} lists the TDCOSMO 2026 milestone sample of lensed quasars, on which we perform our analysis (TDCOSMO 2026, in prep.). In the same table, we list the lens and source redshifts, the telescope or survey with which the wide-field environmental observations were obtained, and the papers in which catalogs and the photometric redshift estimates were first generated from those images. In order to use the number counts technique, it is necessary to have imaging data that both cover a large enough field of view and that are deep enough so that they can be expected to provide a reasonable estimate of the over- or under-density of the line of sight containing the lens.  In TDCOSMO, the standard for the imaging is to acquire, if possible, data over a circular field of 2 arcminutes in radius and to a depth of $i = 24$ or equivalent, based on the analysis presented in \citet{Collett2013}.  For several of the lenses in our sample, the lens system falls within the footprint of a deep survey such as the Dark Energy Survey \citep[DES;][]{Abbott_2021}, which can thus provide the catalogs needed for the number count analysis.  Other lenses in the sample were the subject of targeted observations obtained specifically for the purpose of evaluating the environmental contribution in previous lens environment studies.  For details of the surveys or the previously performed analyses, we refer readers to the references in \cref{tab:lens_sample}.  For several of the lenses in the sample, in particular PG1115+080, RXJ1131$-$1231, and B1608+656, we needed to create new catalogs from targeted observations.  Furthermore, the previous catalogs for HE0435$-$1223 covered a field of view that was too small for our purposes, and thus we needed to recreate the catalogs for that lens as well.  We describe the catalog creation for those four lens systems below.
\subsubsection{Data acquisition and reduction}
The environments of HE0435$-$1223, RXJ1131$-$1231, and B1608+656 were observed with a dedicated program that used the Suprime-Cam instrument \citep{SuprimeCam} on the 8-m Subaru Telescope (Program ID: o14220. PI: Fassnacht).  The data were obtained on 2014 March 1 and consisted of exposures in the $g$, $r$, and $i$ filters for each lens.  The $r$-band data were deep, with total exposure times of 4800~sec per lens, so that they could be used for weak lensing analyses, while the $g$ and $i$-band data had a total exposure times of 600-sec per lens.  The data were calibrated using the SDFRED software \citep[]{Yagi_2002,Ouchi_2004} and then aligned using the Scamp and SWarp packages \citep{terapix, scamp}.  The photometric solutions for HE0435$-$1223,  RXJ1131$-$1231, and B1608+656 were calculated by using an image taken on the same night of a field that overlapped the Sloan Digital Sky Survey (SDSS) footprint.  For more details, see \citet{Rusu_2017}.  

For the PG1115+080 system we used the imaging obtained as part of a monitoring program with the MPIA/ESO 2.2m Telescope \citep{Bonvin2018}.  In all, 95 individual $R_c$-band exposures were combined to create an image with an effective exposure time of 31350~sec.  The photometric calibration was done through comparison to the SDSS catalog of stars detected in the field.

\begin{figure}[!ht]
    \centering
    \includegraphics[width=\linewidth]{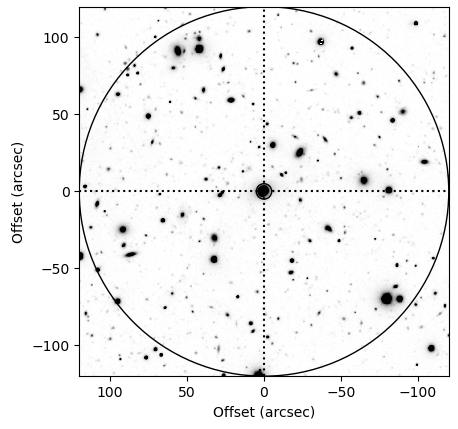}
    \caption{The Subaru/SuprimeCam $i$-band data of the RXJ1131$-$1231 field.  The crosshairs indicate the location of the lens system.  The galaxies that contribute to the number count calculations are located in the annulus between the inner ($r_{\rm in} = 5^{\prime\prime}$) and outer ($r_{\rm out} = 120^{\prime\prime}$) circles.}
    \label{fig:1131_scam_i}
\end{figure}

\begin{figure}[!ht]
    \centering
    \includegraphics[width=\linewidth]{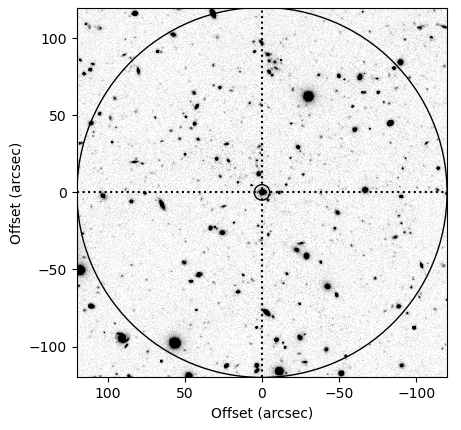}
    \caption{Same as Figure~\ref{fig:1131_scam_i}, but for the $i$-band data of the B1608+656 field.}
    \label{fig:1608_scam_i}
\end{figure}

The images of HE0435$-$1223 and PG1115+080 have been presented in \citet{Rusu_2017} and \citet{Bonvin2018}, respectively.  In Figures~\ref{fig:1131_scam_i} and \ref{fig:1608_scam_i} we show the Subaru SuprimeCam $i$-band images of the RXJ1131$-$1231 and B1608+656 fields.

\subsubsection{Catalog generation}
For all of the fields, the catalogs were generated through the use of a two-pass run of the SExtractor image detection code \citep{BertinArnouts1996}.  The first pass was used to determine the seeing, via an examination of a plot showing the full width at half maximum (FWHM) vs. object magnitude.  In this plot, the stellar locus is clearly visible as long as enough objects have been detected in the field.  The seeing value is then used for separating stars from galaxies in the second pass of SExtractor.  All of the catalogs satisfied the depth and field-of-view criteria for our analysis.
\footnotetext[4]{\hyperlink{http://des.ncsa.illinois.edu/releases/sva1D}{{http://des.ncsa.illinois.edu/releases/sva1D}}} 
\footnotetext[5]{The WIYN Observatory is a joint facility of the NSF's National Optical-Infrared Astronomy Research Laboratory, Indiana University, the University of Wisconsin-Madison, Pennsylvania State University, Purdue University and Princeton University.}
\footnotetext[6]{\hyperlink{https://datalab.noirlab.edu/data/delve$\#$delve-dr3}{https://datalab.noirlab.edu/data/delve$\#$delve-dr3}}
\subsection{Procedural choices}
To estimate the line-of-sight convergence terms for our lens sample, we apply the procedure described in \cref{sec:number_counts}. Unmasked galaxies are counted within an annulus defined by $r\e{in}=5''$ and $r\e{out}=120''$, excluding perturbers whose contribution is explicitly modelled in the upcoming TDCOSMO milestone analysis. By default, we make use of i-band magnitudes with a limiting magnitude of 24, with the exception of PG1115, for which only r-band magnitudes were available, and J1537, for which a limiting magnitude of $i=$23 was needed to ensure completeness \citep{galan_2026}. Our comparison fields are sampled from DES. For each system, we sample a minimum of $\sim 5\times 10^4$ comparison fields, and compare to a minimum of $\sim5\times10^6$ simulation fields.
\subsection{Results}
\begin{figure}[!ht]
    \centering
    \includegraphics[width=\linewidth]{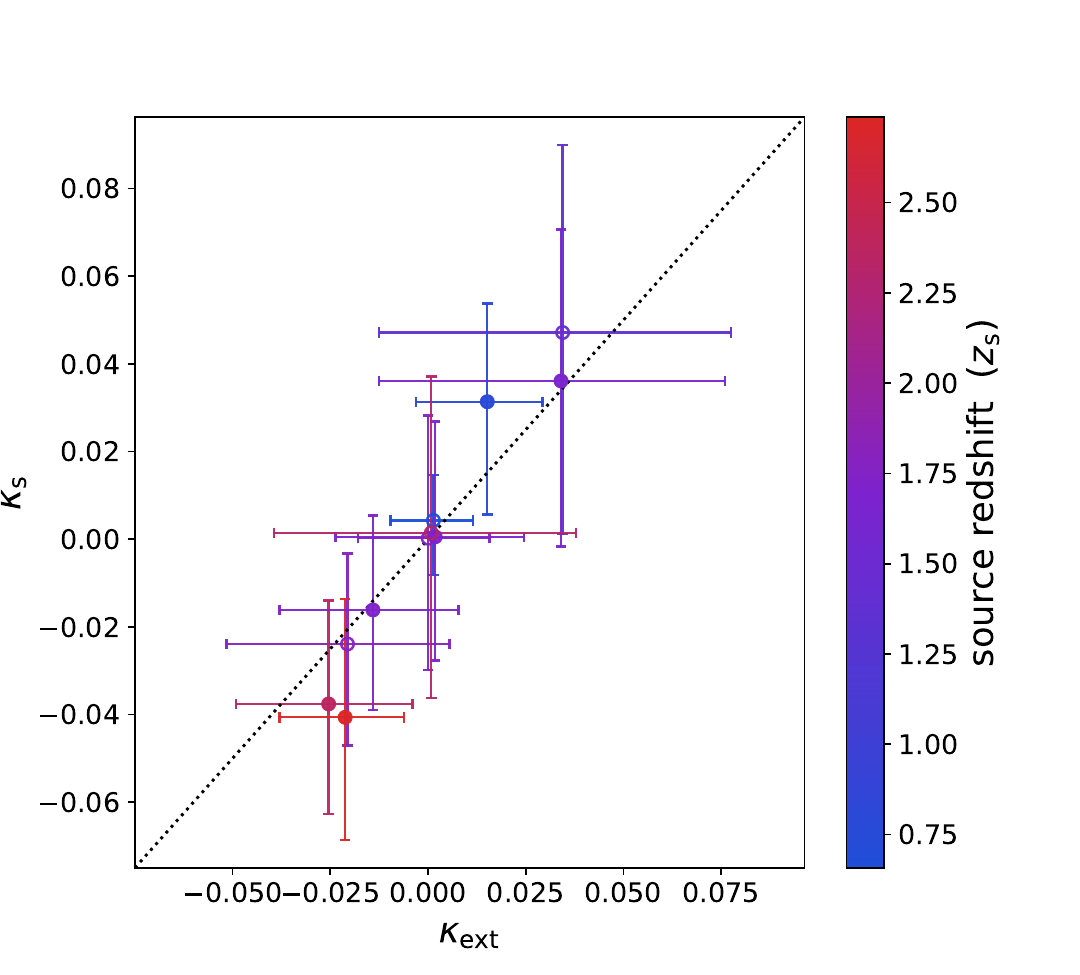}
    \caption{The mean $\kappa\e{ext}$ and $\kappa\e{s}$ values calculated with the method described in \cref{sec:number_counts} for our sample of time delay lenses, with the 16-84th percentile interval plotted as error bars and the source redshift shown as the point colours. Points corresponding to lenses whose environments were constrained without photometric redshifts are shown as open circles. }
    \label{fig:kappa_os_vs_kappa_los_scatter}
\end{figure}
\begin{figure*}[!ht]
    \centering
    \includegraphics[width=\linewidth]{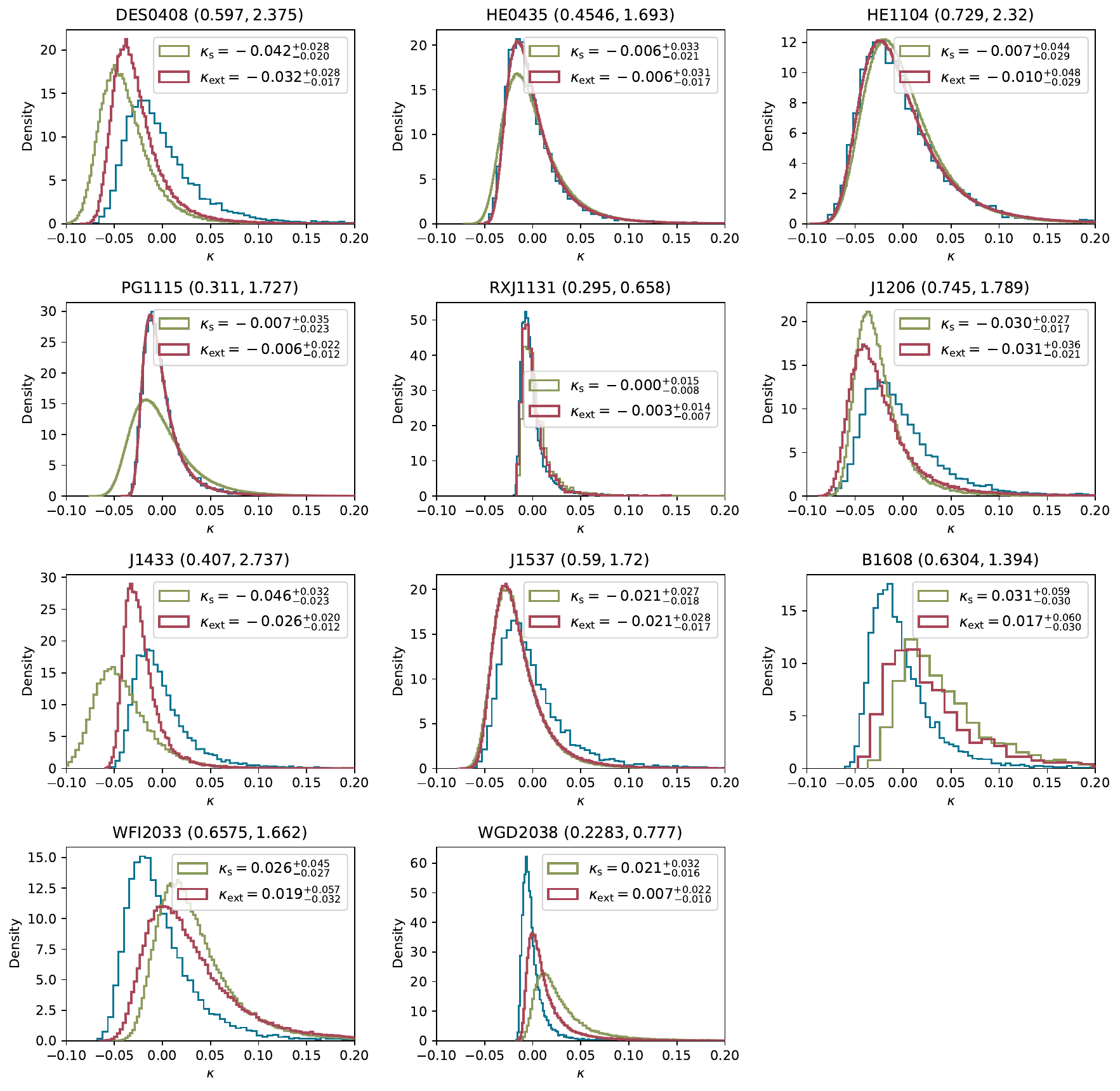}
    \caption{Convergence measurements for our sample of time delay lenses. In red we show the $\kappa\e{ext}$ values obtained using the number counts method described in \cref{sec:number_counts}, our main results. In green, we show the $\kappa\e{s}$ values obtained using the same method. The medians and 1-sigma confidence intervals of these distributions are shown in the legend. In blue, we plot the $\kappa\e{ext}$ distribution obtained for the same lens and source redshift (given in the subplot titles), but sampled randomly from FS2, without using weighted number count information.}
    \label{fig:kappa_os_ext}
\end{figure*}
\begin{figure*}[!ht]
    \centering
    \includegraphics[width=\linewidth]{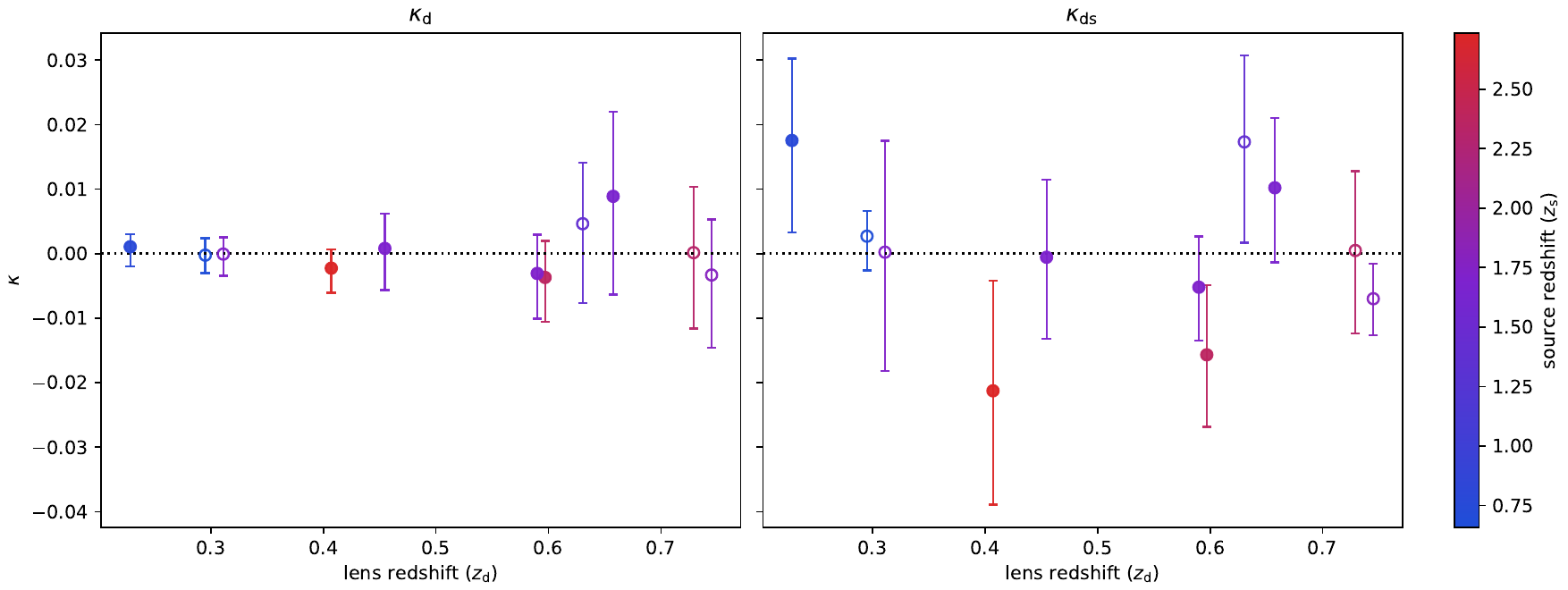}
    \caption{The foreground (left) and background (right) convergence terms, $\kappa\e{d}$ and $\kappa\e{ds}$, plotted as a function of lens redshift, with source redshift indicated by colour. The points indicate the mean of the distributions resulting from the weighted number counts method described in \cref{sec:number_counts}, with error bars corresponding to the 16-84th percentile interval. Points plotted as an open circle as those for which we do not use photometric redshifts in our weighted number count contraints.}
    \label{fig:kappa_ds_od}
\end{figure*}
The histograms of sampled $\kappa\e{ext}$ values for our sample of lenses are shown in \cref{fig:kappa_os_ext}. In the same figures, we also plot the $\kappa\e{s}$ values estimated using the same methods, and the $\kappa\h{unweighted}\e{ext}$ values sampled for our lenses from random lines of sight. We see that, for most of our lenses, the constraints on $\kappa\e{ext}$ are similar to those on $\kappa\e{s}$ and $\kappa\h{unweighted}\e{ext}$. The lines of sight on which B1608, DES2038 and WFI2033 lie appear over-dense compared to random directions in the sky, while J1206, J1433, DES0408 and J1537 are somewhat under-dense. Nonetheless, all are consistent with $\kappa\e{ext}=0$ at 2$\sigma$. 

In \cref{fig:kappa_os_vs_kappa_los_scatter}, we compare the $\kappa\e{ext}$ and $\kappa\e{s}$ values determined for our lens sample. We see that these quantities are highly correlated; from \cref{fig:kappa_os_ext} we see that they may differ by up to $\sim1\sigma$ for an individual lens, but in the majority of cases they are close to identical, and there is no systematic shift in either direction between $\kappa\e{s}$ and $\kappa\e{ext}$.

In \cref{fig:kappa_ds_od}, we show the constraints we obtain on the foreground and background convergences, $\kappa\e{d}$ and $\kappa\e{ds}$ respectively. $\kappa\e{d}$ is more tightly constrained and has a lower scatter than $\kappa\e{ds}$, and these foregrounds do not appear systematically over or under dense, assuaging concerns of biases arising from neglecting this term in previous analyses \citep[see][]{Teodori_2022,Johnson_2024}. However, any definitive conclusion will require a full hierarchical analysis akin to that in \cite{Wells_2024}, and a larger sample of systems. Unsurprisingly, the uncertainty in $\kappa\e{d}$ increases with the lens redshift. $\kappa\e{ds}$ has more scatter, with a standard deviation of mean $\kappa\e{ds}$ values of 0.011 versus just 0.003 for $\kappa\e{d}$, and this is the main driver of differences between $\kappa\e{s}$ and $\kappa\e{ext}$ in this sample, in the cases where these differ. 
\subsection{Comparison to the literature}
\begin{figure*}[!ht]
    \centering
    \includegraphics[width=\linewidth]{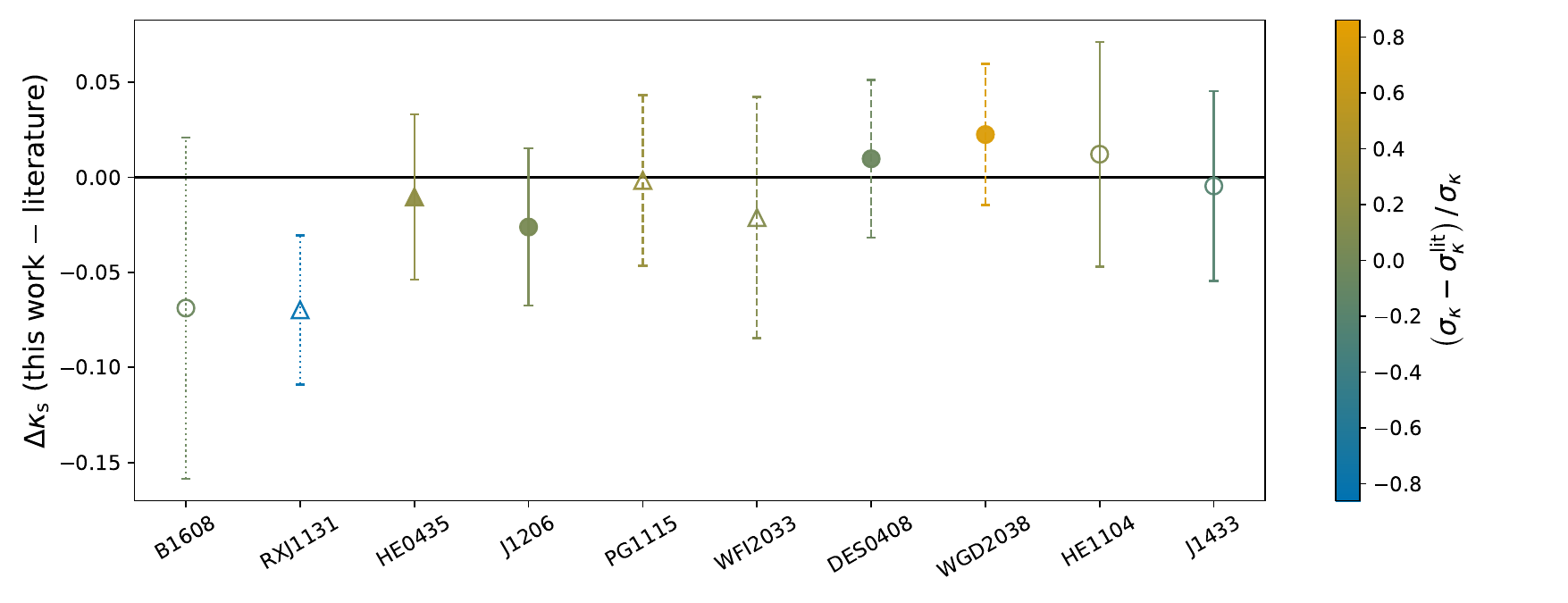}
    \caption{The difference between the $\kappa\e{s}$ values estimated in this work, and selected values reported in the literature. Lenses are ordered by the publication date of the comparison work. Triangular markers correspond to literature values which include the modelled external shear as a constraint, circle markers for those which do not. Filled markers correspond to literature values which include partial or complete spectroscopic redshift information for the environment. The error bars are the quadratic sum of the standard deviations of our distributions, and the uncertainties reported in the literature. The colourbar is used to indicate the relative difference in $1\sigma$ precision between literature values $\left(\sigma_\kappa\h{lit}\right)$ and ours $\left(\sigma_\kappa\right)$ - bluer points are those for which our results are more precise, and yellow where they are less so. Solid error bar lines correspond to literature values derived by considering all galaxies within a 120'' aperture, dotted lines using a 45" aperture, and dashed lines for those which combine weights from both apertures.}
    \label{fig:delta_kappa_os_vs_lit}
\end{figure*}
\begin{figure}[!ht]
    \centering
    \includegraphics[width=\linewidth]{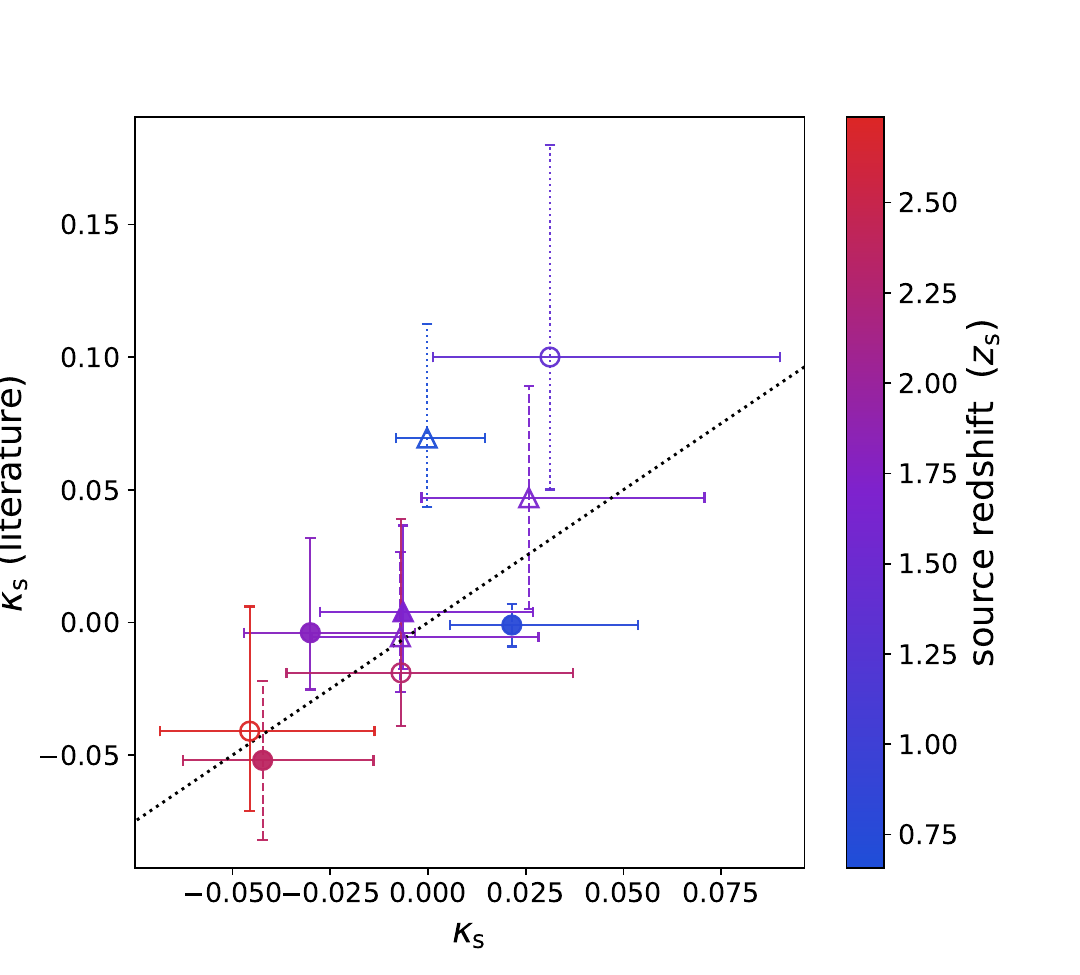}
    \caption{A comparison between our $\kappa\e{s}$ estimates (on the x-axis) and selected values found in other works (on the y-axis) and their reported uncertainties. The line $\kappa\e{s}=\kappa\e{s}$ (literature) is shown as a dotted black line. The markers and linestyles are set as \cref{fig:delta_kappa_os_vs_lit}.}
    \label{fig:kappa_os_vs_kappa_lit_scatter}
\end{figure}
Of the systems studied in this work, estimates of $\kappa\e{ext}$ have been reported for all but J1537-3010, under a range of different assumptions and analysis choices, such as the aperture, limiting magnitude and weighting scheme, and with a diversity of wide-field data, such as the observing telescope and comparison fields, and whether or not spectroscopic redshift data was available. In every case, the Millennium Simulation was used, and these estimates have implicitly or explicitly used $\kappa\e{s}$ as a substitute for $\kappa\e{ext}$, and so we will compare these literature values to our estimates of $\kappa\e{s}$. 

In \cite{Suyu_2010}, the external convergence acting on B1608 was estimated, with the modelled external shear $\gamma\e{ext}$ used as an additional weight. \cite{Suyu_2013} performed a similar analysis for RXJ1131. HE0435 was analysed in \cite{Rusu_2017}, using both $\gamma\e{ext}$ and spectroscopic redshift information. \cite{Birrer2019} analysed J1206, using the same weighting scheme as in our analysis. PG1115 is analysed in \cite{Chen_2019}, WFI2033 in \cite{Rusu2019}, and DES0408 and WGD2038 in \cite{Buckley-Geer_2020}. More recently HE1104 was analysed in \cite{Paic_2026} and J1433 in \cite{2026arXiv260414145S}, using the methods set out in \cite{Wells_2023,Wells_2024}. In some of these papers, multiple $\kappa\e{ext}$ values are reported under different assumptions; in these cases we choose whichever result is presented as fiducial, or whichever result is best constrained when they are presented on equal footing. We refer the reader to these references for more details of each analysis and the choices therein. While some lenses are analysed in multiple other studies, we choose these papers as fiducial estimates used in previous TDCOSMO studies.

Comparisons between these results and ours are shown in \cref{fig:delta_kappa_os_vs_lit,fig:kappa_os_vs_kappa_lit_scatter}. Our results are consistent within $1\sigma$ with each of these estimates except in the case of RXJ1131-1231, and the general identification of lens lines of sight as over or under dense is consistent, as can be seen by the trend in \cref{fig:kappa_os_vs_kappa_lit_scatter}. In the case of RXJ1131, the fiducial value in \cite{Suyu_2013} includes the modelled external shear as a constraint, but also presents a result without this constraint, which we note is consistent with ours to within $1\sigma$. The inclusion of this shear constraint for that system in \cite{Suyu_2013} was in part motivated by the presence of a group of galaxies outside the 120'' aperture used in the analysis, but in the direction of the modelled external shear. While we avoid the use of this shear for the reasons discussed in \cref{sec:external_shear}, it nonetheless remains to be tested whether incorporating the effect of this group, for example by increasing $r_\mathrm{max}$, impacts the estimated convergence.

Certain lens environments have also been studied using weak lensing estimates of the convergence. In \cite{Tihhonova_2018}, these techniques were applied to HE0435-1223, finding a marginally underdense value of $\kappa\e{s} = -0.012^{+0.020}_{-0.013}$, consistent with our estimate at $\approx0.1\sigma$ and a factor of $\sim2$ more precise. In \cite{Tihhonova_2020}, a similar method is used to estimate $\kappa\e{s}=0.11^{+0.06}_{-0.04}$ for B1608+656, in a mild  ($\approx 1.1\sigma$) disagreement with our result with similar precision.

The estimated line-of-sight convergence terms for each of the 11 systems studied in this work can be found in \cref{tab:los_values}. The general consistency between previous estimates of $\kappa\e{s}$ and those obtained in this analysis suggests a degree of robustness to the methodological choices employed, with the possible exception of the inclusion (or lack thereof) of the modelled external shear. Nonetheless, while these $\kappa\e{s}$ estimates are clearly highly correlated with $\kappa\e{ext}$, as is apparent in \cref{fig:kappa_os_ext}, it is $\kappa\e{ext}$, with its foreground and background contributions, with which $H_0$ is degenerate, and which is therefore the quantity which must be used to accurately correct for the impact of the line of sight in time delay cosmography.

%
\begin{table*}[t]
\caption{The estimated $\kappa\e{ext}$ for each of the lenses studied in this work, as well as the observer-to-source term $\kappa\e{s}$, observer-to-lens term $\kappa\e{d}$ and lens-to-source term $\kappa\e{ds}$.}
\label{tab:los_values}
\centering
\small
\renewcommand{\arraystretch}{1.3}
\begin{tabular}{l p{3.0cm} p{3.0cm} p{3.0cm} p{3.0cm} }
\hline\hline
Lens &
$\kappa\e{ext}$ &
$\kappa\e{s}$ &
$\kappa\e{d}$ &
$\kappa\e{ds}$  \\
\hline
DES0408-5354 & $-0.032^{+0.028}_{-0.017}$ & $-0.042^{+0.028}_{-0.020}$ & $-0.006^{+0.008}_{-0.004}$ & $-0.018^{+0.013}_{-0.009}$ \\
HE0435-1223 & $-0.006^{+0.031}_{-0.017}$ & $-0.006^{+0.033}_{-0.021}$ & $-0.002^{+0.008}_{-0.004}$ & $-0.003^{+0.015}_{-0.010}$ \\
HE1104-1805 & $-0.010^{+0.048}_{-0.029}$ & $-0.007^{+0.044}_{-0.029}$ & $-0.004^{+0.014}_{-0.008}$ & $-0.002^{+0.015}_{-0.010}$ \\
PG1115+080 & $-0.006^{+0.022}_{-0.012}$ & $-0.007^{+0.035}_{-0.023}$ & $-0.002^{+0.004}_{-0.002}$ & $-0.004^{+0.022}_{-0.014}$ \\
RXJ1131-1231 & $-0.003^{+0.014}_{-0.007}$ & $-0.000^{+0.015}_{-0.008}$ & $-0.001^{+0.004}_{-0.002}$ & $+0.000^{+0.006}_{-0.003}$ \\
SDSS J1206+4332 & $-0.031^{+0.036}_{-0.021}$ & $-0.030^{+0.027}_{-0.017}$ & $-0.008^{+0.013}_{-0.007}$ & $-0.008^{+0.007}_{-0.005}$ \\
J1433+6007 & $-0.026^{+0.020}_{-0.012}$ & $-0.046^{+0.032}_{-0.023}$ & $-0.004^{+0.004}_{-0.002}$ & $-0.024^{+0.020}_{-0.015}$ \\
J1537-3010 & $-0.021^{+0.028}_{-0.017}$ & $-0.021^{+0.027}_{-0.018}$ & $-0.005^{+0.008}_{-0.005}$ & $-0.007^{+0.010}_{-0.007}$ \\
B1608+656 & $+0.017^{+0.060}_{-0.030}$ & $+0.031^{+0.059}_{-0.030}$ & $-0.000^{+0.014}_{-0.008}$ & $+0.012^{+0.019}_{-0.010}$ \\
WFI2033-4723 & $+0.019^{+0.057}_{-0.032}$ & $+0.026^{+0.045}_{-0.027}$ & $+0.003^{+0.019}_{-0.010}$ & $+0.007^{+0.014}_{-0.009}$ \\
WGD2038-4008 & $+0.007^{+0.022}_{-0.010}$ & $+0.021^{+0.032}_{-0.016}$ & $-0.001^{+0.004}_{-0.001}$ & $+0.012^{+0.018}_{-0.009}$ \\
\hline
\end{tabular}
\end{table*}
\section{Conclusion \label{sec:conclusion}}
In this work, we have presented developments to the weighted number counts method to estimate the external convergence of strong gravitational lenses. Our main advancement compared to previous methods is the inclusion of the foreground and background convergence terms, which we have argued makes the correction for the external mass-sheet degeneracy more accurate. By replacing the Millennium Simulation products used in previous analyses with the \textit{Euclid} Flagship Simulation, we have exploited an improved particle mass and angular and redshift resolution, a dramatic increase in simulation volume and hence the statistical power of the method, a greater degree of homogenisation in the estimation of quantities such as filter-specific magnitudes, photometric redshifts and weak lensing convergence, an updated cosmological model, and improvements in halo population models. Finally, we have argued that, for homogeneity in the analysis and to alleviate potential systematics, the external shear should not be used as an additional weight by which simulated lines of sight are selected.

We have applied these methods to the TDCOSMO 2026 milestone sample of 11 lensed quasars used in time delay cosmography. These results constitute a standard set of external convergence posteriors which are used within the 2026 milestone hierarchical analysis, as well as individiually within \cite{williams_2026,galan_2026}. While some of these lenses are individually over or under dense, this does not seem to be systematic in either direction, and the mean external and observer-source convergences of the sample are very close to zero. Comparing our $\kappa\e{s}$ results to values reported in previous studies using weighted number counts methods, our estimates are consistent to $1\sigma$ for all but one lens, RXJ1131, for which the difference appears driven by our exclusion of the modelled external shear as a weight. 

Our method enables the foreground and background convergence terms (as well as the usual observer to source term) to be estimated individually. For our sample of lenses, we find that the foreground convergence terms are typically small, consistent with zero with $1\sigma$ errors which are well below one percent for the majority of our lenses, and corresponding to foregrounds which do not appear to be systematically over or under dense, minimising the bias arising from neglecting this term in velocity dispersion constraints of the mass-sheet parameter \citep{Teodori_2022,Johnson_2024}. These methods will allow such biases to be removed completely in future studies. 

The methods described in this paper can be used to estimate the convergence between arbitrary redshifts on a given line of sight, which is essential for precision cosmology with double-plane lenses, where line of sight corrections are both needed and poorly approximated by a single observer to source term \citep{Johnson_2025,Johnson_2026,Teodori_2026}. It is therefore for these systems where we expect the value of these methods to be most pronounced. Having access to the foreground and background convergence individually will also enable more detailed studies of the role played by the line of sight in the selection function of strong gravitational lenses. 

Our sample of lenses, while large enough for competitive constraints from time delay cosmography, is too small for a meaningful population-level study of lens environments and selection effects. In future, we aim to apply this analysis both to previously studied lens samples, and to the host of new systems being discovered by ongoing stage-IV surveys such as \textit{Euclid} \citep{Walmsley_2025}, and to perform the full hierarchical inference necessary to identify any systematic trends hinted at by the data.

The volume and detail of the \textit{Euclid} Flagship Simulation significantly reduces the statistical limitations of the weighted number count method. As expected and as is apparent in our results, the summary statistics used to weight the sampling of simulated lines of sight do not fully capture the correlations present between visible structures and the projected external mass density on a given line of sight, and so in future we hope to further optimise these weights and hence further reduce the uncertainty in the external convergence estimates. 


\begin{acknowledgements}

We would like to give particular thanks to Pau Tallada and Jorge Carretero from the \textit{Euclid} Flagship Simulation team for their extensive help in our use of the simulation products, and to Dominique Sluse for reviewing a draft of this work, and for several very helpful discussions. DJ acknowledges support by the First Rand Foundation, South Africa,
and the Centre National de la Recherche Scientifique of France.  This work has made use of CosmoHub, developed by PIC (maintained by IFAE and CIEMAT) in collaboration with ICE-CSIC. It received funding from the Spanish government (grant EQC2021-007479-P funded by MCIN/AEI/10.13039/501100011033), the EU NextGeneration/PRTR (PRTR-C17.I1), and the Generalitat de Catalunya. PF acknowledges
support from the French Agence Nationale de la Recherche through the ELROND project
(ANR-23-CE31-0002). MM acknowledges support by the SNSF (Swiss National Science Foundation) through Ambizione grant PZ00P2\_223738. 

\end{acknowledgements}

\bibliographystyle{aa}
\bibliography{main}


\begin{appendix}
%

\section{Discrete integration}
\label{app:discrete_integration}
\begin{figure}[!ht]
    \centering
    \includegraphics[width=\linewidth]{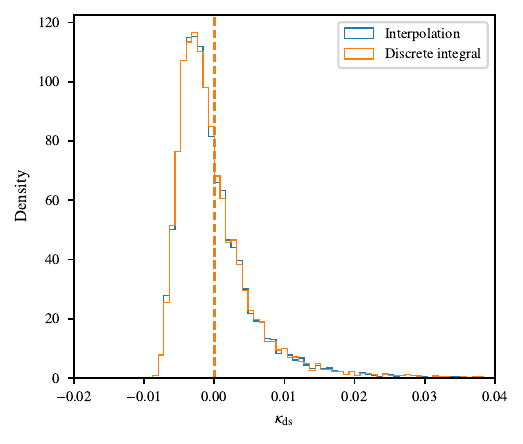}
    \caption{The $\kappa\e{ds}$ histogram for a lens and source at redshifts $z\e{d}=0.5$, $z\e{s}=1.0$, sampled from FS2 without number count weights. In blue, $\kappa\e{ds}$ is calculated from a smooth interpolation of $\kappa(\chi)$ and integrated with \texttt{quad} \cite{Virtanen_2020}. In orange, $\kappa\e{ds}$ is approximated using \cref{eq:kappa_ds_approx}.}
    \label{fig:interpolation_vs_discrete}
\end{figure}
In this section, we detail the numerical methods used to speed up the calculation of \cref{eq:kappa_ij_from_kappa_s}. Defining $y_n\equiv\chi_n\kappa(\chi_n)$, we can estimate the second derivative with respect to $\chi$ as
\begin{equation}
    y''(\chi_n) \approx \frac{2}{\chi_{n+1}-\chi_{n-1}}\pac{\frac{y_{n+1}-y_n}{\chi_{n+1}-\chi_n}-\frac{y_n-y_{n-1}}{\chi_n-\chi_{n-1}}}.
\end{equation}
The weight function $W\e{ds}(\chi_n)$ is evaluated for all $\chi_i < \chi_n < \chi_j$, and set to 0 outside these bounds. A Voronoi-like integration cell is assigned to each element, with a cell width defined between the midpoints of neighbouring shells, clipping any region which lies outside the integration bounds, i.e.
\begin{equation}
    \Delta\chi_n = \mathrm{min}\pa{\frac{\chi_{n-1}+\chi_n}{2},\chi_j}-\mathrm{max}\pa{\frac{\chi_n+\chi_{n+1}}{2},\chi_j}.
\end{equation}
We can then estimate the integral with a Riemann sum of these quantities over the available redshift planes
\begin{equation}
    \kappa\e{ds} \approx \sum^{N-2}_{n=1}\pac{\frac{y''(\chi_n)}{\chi_n}W\e{ds}(\chi_n)}\Delta\chi_n.
    \label{eq:kappa_ds_approx}
\end{equation}
To test the reliability of this method, \cref{fig:interpolation_vs_discrete} compares the $\kappa\e{ds}$ histograms obtained using i. an interpolated $\kappa(\chi)$ and a numerical integral, and ii. the discrete method described above, with randomly sampled lines of sight for a lens at $z=0.5$ and a source at $z=1$. The resulting distributions are almost identical, with a difference in mean value $\approx5\times10^{-6}$ and standard deviation $\approx3\times10^{-5}$.
\end{appendix}

\end{document}